\documentclass[10pt,twocolumn,
 superscriptaddress,
 longbibliography,
 aps,showkeys, floatfix]{revtex4-2}
 
\usepackage{float} 
\usepackage{setspace}
\usepackage{enumitem}
\usepackage{ dsfont }
\usepackage[pdftex]{hyperref,color,graphicx}
\usepackage{amsfonts,amssymb,amsmath}
\usepackage[separate-uncertainty=false]{siunitx}
\usepackage[english]{babel}
\usepackage[utf8]{inputenc}
\usepackage[T1]{fontenc}
\usepackage[normalem]{ulem}
\usepackage[toc,page]{appendix}
\usepackage[space]{grffile}
\usepackage{placeins}
\usepackage{latexsym}
\usepackage{textcomp}
\usepackage{longtable}
\usepackage{multirow,booktabs}
\usepackage{url}
\hypersetup{colorlinks=false,pdfborder={0 0 0}}
\usepackage{dcolumn}
\usepackage{bm}
\usepackage{dsfont}
\usepackage{upgreek}
\usepackage{cancel}
\usepackage{braket}
\usepackage{color}
\usepackage[normalem]{ulem}

\setlist[itemize]{topsep=.1em,itemsep=.1em,parsep=0em,partopsep=0em}
\setlist*[itemize]{first=\vspace{\baselineskip}\setstretch{1}\vspace{-\baselineskip}}    
\newcommand{\be}{\begin{equation}}
\newcommand{\ee}{\end{equation}}
\newcommand{\bea}{\begin{equation} \begin{aligned}}
\newcommand{\eea}{\end{aligned} \end{equation}}
\newcommand{\sig}{\sigma}

\newcommand{\Gq}{\Gamma_q}
\newcommand{\Ga}{\Gamma_a}
\newcommand{\Gb}{\Gamma_b}

\newcommand{\gq}{\gamma_q}
\newcommand{\ga}{\gamma_a}
\newcommand{\gb}{\gamma_b}

\begin{document}


\title{Detecting entanglement between quantum emitters using directional emission}
\author{Ivan Saychenko}
\affiliation{Department of Mathematical, Physical and Computer Sciences, University of Parma, Parco Area delle Scienze 7/A, 43124, Parma, Italy}	
\affiliation{INFN, Sezione di Milano Bicocca, Gruppo Collegato di Parma, Parco Area delle Scienze 7/A, 43124 Parma, Italy}
\email{ivan.saychenko@unipr.it}

\author{Robert Wei\ss}
\affiliation{Institut f\"ur theoretische Physik, Universit\"at Heidelberg, Philosophenweg 12, 69120 Heidelberg, Germany}	
\email{robert.weiss@stud.uni-heidelberg.de}

\author{Scott Parkins}
\affiliation{Dodd-Walls Centre for Photonic and Quantum Technologies, Auckland, New Zealand}
\affiliation{Department of Physics, University of Auckland, Auckland 1010, New Zealand}	
\email{s.parkins@auckland.ac.nz}

\author{Rita Veilande}
\affiliation{Institute of Atomic Physics and Spectroscopy, University of Latvia, Raina boul. 19, Riga, LV-1586, Latvia}	
\email{rita.veilande@lu.lv}

\author{Mark Sadgrove}
\affiliation{Department of Physics, Faculty of Science, Tokyo University of Science, 1-3 Kagurazaka, Shinjuku-ku, Tokyo 162-8601, Japan}
\email{mark.sadgrove@rs.tus.ac.jp}

\author{Sandro Wimberger}
\affiliation{Department of Mathematical, Physical and Computer Sciences, University of Parma, Parco Area delle Scienze 7/A, 43124, Parma, Italy}
\affiliation{INFN, Sezione di Milano Bicocca, Gruppo Collegato di Parma, Parco Area delle Scienze 7/A, 43124 Parma, Italy}
\affiliation{National Quantum Science and Technology Institute, Spoke 1, University of Parma, 43124 Parma, Italy}
\email{sandromarcel.wimberger@unipr.it}

\date{\today}

\begin{abstract}
Recently, it was shown that quantum interference in a system containing a polarized and unpolarized emitter can allow directional emission of photons into a circulating cavity.
Here, we ask whether high directionality of photon emission in this system implies a high degree of quantum correlation between the two emitters.
We show that the answer is a qualified "yes", with photon emission directionality and emitter-emitter entanglement showing a monotonic relationship over a broad parameter range. The relationship
only breaks down in the limit of perfect directionality. Furthermore, under reasonable assumptions for experimental parameters and stability, we show that the
statistics of measured directionality allow a reliable estimate of the concurrence. This result implies that directionality of photon emission in the state preparation stage can
be used to determine the entanglement between the emitters, with potential applications to more generic cases including quantum networks.
\end{abstract}


\maketitle

\section{Introduction}

In recent years, directional coupling between quantum emitters and propagating optical modes via evanescent fields has attracted attention due to its potential for realizing novel quantum optical devices~\cite{petersen2014chiral,scheucher2016quantum,sayrin2015nanophotonic,pucher2022atomic,rosenblum2016extraction,pichler2015quantum}. In standard cavity QED setups, an emitter typically couples equally to left and right propagating modes due to symmetry. However, the
spin-momentum locking~\cite{bliokh2015spin,bliokh2014extraordinary,van2016universal,petersen2014chiral} found in evanescent fields allows this symmetry to be broken by matching the handedness of the emitter's dipole moment to that of the field polarization.
Although this effect has been convincingly demonstrated using cold atoms and solid state emitters at cryogenic temperatures, the requirement of a well-defined, approximately circular dipole moment for the emitter
greatly reduces the applications of the technique. For example, the polarization of photons emitted from room temperature solid state quantum emitters is typically random~\cite{abe2017dynamically}, precluding the possibility of directional coupling.

Recently, it was shown that the directional coupling effect can essentially be transferred from a polarized, three level emitter (henceforth referred to as "the atom") to an unpolarized one (henceforth "the quantum dot" (QD))
using cavity QED techniques~\cite{Ostrowski_2022}. In that work,
the amplitude for coupling to one of the modes of a circulating cavity was shown to be cancelled due to quantum interference, leading to effective directional emission from the unpolarized emitter
into the cavity. A natural question then arises: Does a large degree of directionality correlate with a large degree of entanglement between the two emitters?

To answer this question, it is necessary to consider
a steady state of the system. In the absence of spontaneous emission, the steady states are the system ground state and a cavity dark state (CDS) in which emission into either cavity mode is prevented by interference, leading to decoupling of the atom-QD system from the cavity. We therefore choose to consider the relationship between entanglement of the cavity dark state and the directionality of emission.

A priori, it is not intuitively clear that directionality of emission is correlated with entanglement of the dark state. The extreme cases suggest not: for non-directional emission, the QD and atom product states
have equal amplitude, which, combined with the conditions required for large directionality gives a nearly separable state (we elaborate on this in the main text). On the other hand, maximum directionality is achieved by removing the third atomic level which destroys the dark state itself, and thus its concurrence. However, between these extreme cases lies the region of interest, for which the relation between directional emission and entanglement is not obvious.

\begin{figure}[tb!]  
	\centering
		\includegraphics[width=0.95\linewidth]{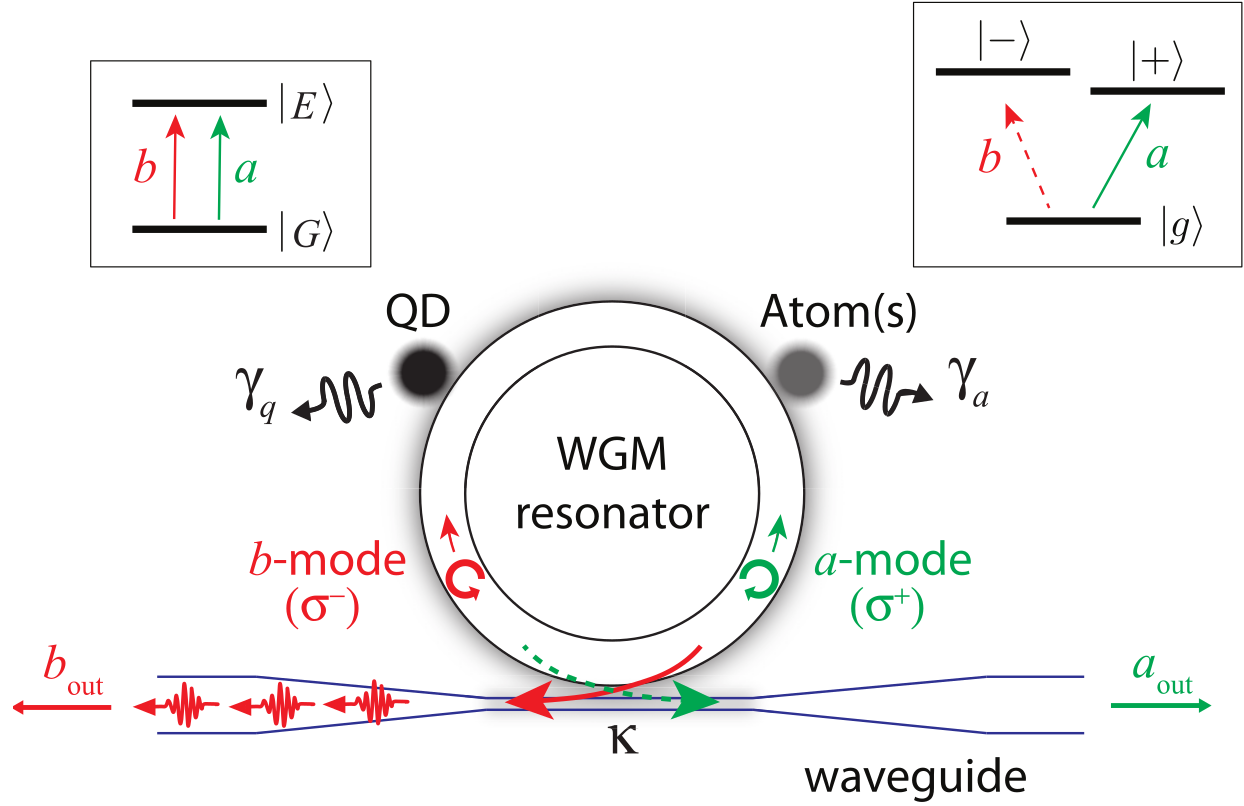}
\caption{Illustration of the system under consideration. A two-level quantum dot (QD) and a V-type three-level atom are coupled to the same whispering-gallery-mode resonator. We label the counterclockwise circulating mode (polarization $\sigma^+$) of the resonator $a$ and the clockwise circulating mode (polarization $\sigma^-$) $b$. The depicted decay channels are QD spontaneous emission at rate $\gamma_q$, atomic spontaneous emission at rate $\gamma_a$, and decay from each cavity mode at rate $\kappa$ into a waveguide. Figure adapted from Ostrowski {\it et al.},~\cite{Ostrowski_2022}.
}
		\label{fig:1}
	\end{figure}

Here, we will show that for a broad parameter regime, there is a monotonic relation between the directionality of emission and the entanglement of the dark state. Our work is an extension of Ref.~\cite{Ostrowski_2022}
in which the atom-QD correlations were not considered. As shown in Fig. \ref{fig:1}, we consider a circular cavity coupled to a quantum emitter, e.g., a quantum dot (QD), and an optical nanofiber. The QD can emit a photon into the cavity, with a random circulation (mode), clockwise or counter-clockwise. This photon will decay into the fiber, propagating  to the left or the right, respectively, see Fig. \ref{fig:1}.

While in the absence of the atom all photons will eventually be emitted into the fiber, the presence of the atom allows for both directionality of photon emission, and for the system to be trapped in the so-called cavity dark state, where no photons are emitted. This state is composed of the product states of the QD and the atom, and, in general, has non-zero entanglement. The relatively low probability of preparing this state means that in principle, many cycles of excitation and directional emission can occur before it is achieved.
The motivation of the present work is to learn the entanglement of the CDS by measuring photons emitted before the system "turns dark" (i.e. falls into the CDS). Practically, it is difficult to measure the entanglement of the CDS directly, and so the use of directionality to determine the entanglement of the CDS is potentially useful.

Here, we use the concurrence as our measure of entanglement. The concurrence is expressed in terms of the couplings between the different parts of the system, which values may be difficult to obtain through direct measurement.
However, since the directionality can also be expressed in terms of these coupling strengths it is possible to obtain the value of the concurrence by measuring only the directionality. Our contribution is to quantify the correlation between concurrence and directionality, allowing the degree of entanglement of the CDS to be ascertained through an experimentally accessible quantity.

The paper is organized as follows:  In Section~\ref{sec-2}, we review the theoretical treatment of the system, and relax some assumptions made in Ref.~\cite{Ostrowski_2022} to find the directionality in the case of a three level atom.
In Section~\ref{sec:ent}, we calculate the concurrence of the CDS, under the assumption that it is long lived relative to the cavity lifetime. In Section~\ref{sec:exp}, we analyse an experimental protocol in which the concurrence is inferred directly from measurements of the directionality, and show that the concurrence of the CDS can be reliably estimated. Finally, we offer a discussion of ideas for future directions along with a conclusion in Section~\ref{sec:concl}.

\section{Theoretical model}
\label{sec-2}
		
Our system displayed in Fig. \ref{fig:1} is composed of: {\em (i)} a QD emitter with a ground state $|G\rangle$ and an excited state $|E\rangle$; {\em (ii)} a three-level atom with the ground state $|g\rangle$ and two excited states $|+\rangle$ and $|-\rangle$; and {\em (iii)} a circular cavity with two orthogonal modes: the counterclockwise-propagating $a$ and the clockwise-propagating $b$. We denote the population in the two modes $|n\rangle_a$ and $|m\rangle_b$, where $n$ and $m$ are the number of photons in the respective mode. For the reader's convenience, our notation closely follows Ref.~\cite{Ostrowski_2022}. The two cavity modes are coupled to the QD with strength $g_q$, the $a$ mode is coupled to the $|+\rangle$ state of the atom with strength $g_a$ and the $b$ mode to the $|-\rangle$ with strength $g_b$. The QD and the atom can be de-excited via spontaneous emission, with the decay rate $\gamma_q$, $\gamma_a$ and $\gamma_b$, respectively for the transitions $|E\rangle \rightarrow |G\rangle$, $|+\rangle \rightarrow |g\rangle$ and $|-\rangle \rightarrow |g\rangle$. The cavity modes decay into the waveguide with the rate $\kappa$. Finally, the QD is continuously driven by a laser with Rabi frequency $\Omega$. Within this model, the Hilbert space $\mathcal{H}$ is
		\bea
			\mathcal{H} &= \text{span} \{|\text{QD}\rangle \otimes |\text{atom}\rangle \otimes |n\rangle_a \otimes |m\rangle_b \} \\
			& \equiv \text{span}\{|\text{QD},\text{atom}, n, m\rangle \},
			\label{eq:Hilbert}
		\eea
where $\text{QD} \in \{G,E\}$, $\text{atom} \in \{g,+,-\}$ and $n \in \mathds{N}$. The Hamiltonian of the system is given within the rotating wave approximation by
\begin{multline} \label{eq:Ham}
			\hat H =
			\Delta_q \hat \sig_{q+} \hat \sig_{q-} + \Delta_a \hat \sig_{a+} \hat \sig_{a-} + \Delta_b \hat \sig_{b+} \hat \sig_{b-} \\
			+ \Big( g_q \sig_{q+}(\hat a + \hat b) + g_a \hat \sig_{a+} \hat a + g_b \hat \sig_{b+} \hat b + \text{H.c.} \Big) \\
			+ \Omega (\hat \sig_{q+} + \hat \sig_{q-}) +\Delta_c(\hat a^\dag \hat a + \hat b^\dag \hat b),
\end{multline}
where H.c. denotes the Hermitian conjugate and with the ladder operators:
	\bea
		\hat \sigma_{q-} &= |G\rangle \langle E| \quad \text{and} \quad \hat \sigma_{q+} = \hat \sigma_{q-} ^\dag \\
		\hat \sigma_{a-} &= |g\rangle \langle +| \quad \text{and} \quad \hat \sigma_{a+} = \hat \sigma_{a-} ^\dag \\
		\hat \sigma_{b-} &= |g\rangle \langle -| \quad \text{and} \quad \hat \sigma_{b+} = \hat \sigma_{b-} ^\dag \\
		\hat a &= |0\rangle_a \langle 1|_a \,\, , \,\,
		\hat b = |0\rangle_b \langle 1|_b.
	\eea

This Hamiltonian is written in a frame rotating at rate $\omega_L$, corresponding to the frequency of the laser driving the QD with strength $\Omega$. The detunings in $\hat H$ are defined as the difference between $\omega_L$ and the transition frequencies between the ground and excited states $\omega_q$ for the QD and  $\omega_\pm$ for the atom or the resonance frequency of the cavity $\omega_c$ by $\Delta_q \equiv \omega_q - \omega_L$, $\Delta_a \equiv \omega_+ - \omega_L$, $\Delta_b \equiv \omega_- - \omega_L$ and $\Delta_c \equiv \omega_c - \omega_L$.

In principle, the system has to be described by a Master equation, see ref. \cite{Ostrowski_2022}, that takes the decay of the cavity, the QD and the coupling into the nanofiber into account. Assuming no driving $\Omega=0$ and that the system is initialized with the QD being in its excited state, the atom in its ground state, and the two cavity modes in the vacuum state,
the Hamiltonian reduces to the first two lines in Eq. \eqref{eq:Ham}.

The system is then restricted to one of the following situations: it can either contain one excitation ($|E,g,0,0\rangle$, $|G,+,0,0\rangle$, $|G,-,0,0\rangle$, $|G,g,1,0\rangle$, $|G,g,0,1\rangle$), or be in the ground state $|G_0\rangle = |G,g,0,0\rangle$. For times $t>0$, the system either emits a photon with the probability $P$ and falls into $|G_0\rangle$, or remains in the following pure state $|\bar \psi \rangle$ with the probability $1-P$:

\bea \label{eq:psi_bar}
		 |\bar \psi(t)\rangle =\ & Q(t) |E,g,0,0\rangle \\
		 & + A(t) |G,+,0,0\rangle + B(t) |G,-,0,0\rangle \\
		 & + \alpha(t) |G,g,1,0\rangle + \beta(t) |G,g,0,1\rangle.
\eea

Here $Q(t)$ is the probability amplitude for the QD being excited, $A(t)$ and $B(t)$, for the atom being in the excited state $|+\rangle$ and $|-\rangle$, and $\alpha(t)$ and $\beta(t)$ are the excitation amplitudes of the cavity modes $a$ and $b$, respectively. The evolution equation of the pure state $|\bar \psi \rangle$ is given by 		
\be \label{Sch_eq}
			i \frac{d\ket{\bar \psi}}{dt} = 
			\hat H_{NH}
			\ket{\bar \psi},
\ee
with the non-Hermitian Hamiltonian $\hat H_{NH}$ that includes the spontaneous emission of the QD and the atom
\begin{multline} \label{eq:H_NM}
			\hat H_{NH} \equiv  \Big( g_q \sig_{q+}(\hat a + \hat b) + g_a \hat \sig_{a+} \hat a + g_b \hat \sig_{b+} \hat b + \text{H.c.} \Big) \\
			 - i\kappa(\hat a^\dag \hat a + \hat b^\dag \hat b) 
			-i\frac{\gq}{2} \hat \sig_{q+} \hat \sig_{q-}
     	-i\frac{\ga}{2} \hat \sig_{a+} \hat \sig_{a-}
     	-i\frac{\gb}{2} \hat \sig_{b+} \hat \sig_{b-},
\end{multline}

Following 	ref. \cite{Ostrowski_2022}, one finally arrives at the evolution equations for the amplitudes
\bea \label{eq:syst}
			\dot{Q}(t)      = & -(\gq/2 + i\Delta_q)Q(t) - ig_q\alpha(t) - ig_q\beta(t)\\
			\dot{A}(t)      = & -(\ga/2 + i\Delta_a)A(t) - ig_a\alpha(t)\\
			\dot{B}(t)      = & -(\gb/2 + i\Delta_b)B(t) - ig_b\beta(t) \\
			\dot{\alpha}(t) = & -\kappa \alpha(t) - i g_q Q(t) - i g_a A(t)\\
			\dot{\beta}(t)  = & -\kappa \beta(t)  - i g_q Q(t) - i g_b B(t).
\eea
	
\subsection{Adiabatic elimination of the cavity modes}
	\label{subsec:adiab}
	
From now on, we will restrict our analysis to the regime of a bad cavity and a large cooperativity, i.e., at $\gamma_{q,a,b} \ll g_{q,a,b} \ll \kappa$. In this regime, the decay of the cavity modes is the fastest process, while the spontaneous emissions of the QD and the atom are the slowest ones. Thus, a photon created in the cavity is "instantly" emitted through the cavity decay channel of the coupled waveguide, while the decays of the QD and atom will occur "after a long time", compared to the characteristic times of the evolution.

Analysis of the system can be taken further by adiabatically eliminating the cavity modes in Eq. \eqref{eq:syst} from the dynamics. This is achieved by setting the time derivatives of $\alpha$ and $\beta$ to zero. To achieve closed form expressions, we further assume that all the transitions within the QD and the atom are resonant with the cavity frequency ($\Delta_{q,a,b}=0$) and we neglect the spontaneous emission of the QD and the atom ($\gamma_{q,a,b} = 0$).
This derivation was already performed in \cite{Ostrowski_2022} in the ideal case limit, i.e., where $g_b = 0$. In this limit, the dynamics of the probability amplitude of the atomic state $|-\rangle$ is decoupled from the rest of the system and does not interfere with the cavity $b$ mode. However, for complete decoupling of the $|-\rangle$ state, no cavity dark state exists, as the quantum dot can always couple to the $b$ mode. This makes the analytical expressions developed in \cite{Ostrowski_2022} unsuitable for a study of the CDS.

Here, we redo the derivation without the additional assumption on $g_b$. This extension of the procedure from  \cite{Ostrowski_2022} gives the solutions for the dot and atom amplitudes of the state $ |\bar \psi(t)\rangle$:
\bea 
			Q(t) = &  \frac{1}{2} \Bigg(1 + \frac{\Ga + \Gb - \Gq}{\sqrt{(\Ga - \Gb)^2 + \Gq^2}} \Bigg) e^{\lambda_+t}  \\
			       & +\frac{1}{2} \Bigg(1 - \frac{\Ga + \Gb - \Gamma_2}{\sqrt{(\Ga - \Gb)^2 + \Gq^2}} \Bigg) e^{\lambda_-t},  \\
			A(t) = & \sqrt{\frac{\Ga \Gq}{2\Big((\Ga - \Gb)^2 + \Gq^2 \Big)}}(e^{\lambda_-t}-e^{\lambda_+t}),  \\ 
			B(t) = & \sqrt{\frac{\Gb \Gq}{2\Big((\Ga - \Gb)^2 + \Gq^2 \Big)}}(e^{\lambda_-t}-e^{\lambda_+t}). 
			\label{eq:proba_amp}
\eea
From these we obtain for the cavity modes
\bea \label{eq:cavity_modes_amp}
			\alpha(t) = &  -\frac{ig_q}{2\kappa} \Bigg( \Bigg[1 - \frac{g_a^2 - g_b^2 + 2g_q^2}{\sqrt{(g_a^2 - g_b^2)^2 + 4g_q^4}} \Bigg]e^{\lambda_+ t} \\
			&+ \Bigg[1 + \frac{g_a^2 - g_b^2 + 2g_q^2}{\sqrt{(g_a^2 - g_b^2)^2 + 4g_q^4}} \Bigg]e^{\lambda_- t} \Bigg) \\
			\beta(t) = &  -\frac{ig_q}{2\kappa} \Bigg( \Bigg[1 + \frac{g_a^2 - g_b^2 - 2g_q^2}{\sqrt{(g_a^2 - g_b^2)^2 + 4g_q^4}} \Bigg]e^{\lambda_+ t} \\
			&+ \Bigg[1 - \frac{g_a^2 - g_b^2 - 2g_q^2}{\sqrt{(g_a^2 - g_b^2)^2 + 4g_q^4}} \Bigg]e^{\lambda_- t} \Bigg).
\eea		
Above we used the definitions
\be
	\Gq \equiv \frac{2g^2_q}{\kappa},\quad \Gamma_{a,b} \equiv \frac{g^2_{a,b}}{\kappa},
\ee
and
\be
	\lambda_\pm \equiv \frac{-(\Ga+\Gb+\Gq) \pm \sqrt{(\Ga - \Gb)^2 + \Gq^2} }{2}.
\ee

\subsection{Directionality of photon emission}
\label{subsec:dir}
		
The (normalized) asymmetry between the probability for a photon to be emitted into the $a$ mode or the $b$ mode is typically referred to as the directionality, as the different circulation directions of the two modes lead to them
outcoupling to opposite propagation directions.
			
The probability of a certain cavity mode emitting a photon at some time during the temporal evolution is obtained by integrating the cavity mode amplitudes Eq. \eqref{eq:cavity_modes_amp} from the initial time $t_\text{initial}=0$ to the final time, formally $t_\text{final} \rightarrow \infty$. In practice, we have to ensure that $t_\text{final}$ is large compared to the characteristic decay time of the cavity, i.e.,  $t_\text{final} \gg 1/\kappa$. Then the probability of a certain mode emitting a photon is
\bea \label{eq:proba}
			P_a \equiv & 2\kappa \int_0 ^{\infty}dt |\langle 1|_a \psi(t)\rangle|^2 = 2\kappa \int_0 ^{\infty}dt |\alpha(t)|^2 \\
			P_b \equiv & 2\kappa \int_0 ^{\infty}dt |\langle 1|_b \psi(t)\rangle|^2 = 2\kappa \int_0 ^{\infty}dt |\beta(t)|^2,
\eea
where the time is expressed in units of $1/\kappa$. 
The directionality is then defined as
\be \label{eq:D_def}
			D \equiv \frac{P_b - P_a}{P_b + P_a}.
\ee
The probabilities $P_a$ and $P_b$ can be rewritten using Eq. \eqref{eq:cavity_modes_amp} in Eq. \eqref{eq:proba} in terms of the couplings strengths and the decay rate
		\bea \label{eq:proba-2}
			P_a = & \frac{4}{\lambda_+\lambda_-}\frac{\Gq(\Ga+\Gb)(\Gq + 2\Gb)}{\Gq+\Ga+\Gb} \\
			P_b = & \frac{4}{\lambda_+\lambda_-}\frac{\Gq(\Ga+\Gb)(\Gq + 2\Ga)}{\Gq+\Ga+\Gb},
		\eea
giving finally

\be \label{eq:D_Gamma}
			D = \frac{\Ga -\Gb}{\Ga + \Gb + \Gq} = \frac{g_a^2-g_b^2}{g_a^2+g_b^2+2g_q^2}.
\ee

This formula reveals the need for a hierarchy of coupling strengths to achieve directionality close to 1, as discussed in \cite{Ostrowski_2022}. In particular, for predominant emission into the $b$-mode, it is necessary that
$g_q, g_b\ll g_a$ to achieve a near unity directionality. We will typically require at least one of these conditions to be true, even as we investigate how the entanglement between emitters behaves in the low directionality limit.
Equation~\eqref{eq:D_Gamma} also shows that we can not obtain perfectly directional photon emission (i.e. $D=1$), since even if we set $g_b=0$, the necessity of $g_q>0$ guarantees that perfect directionality is not achieved.
Figure \ref{fig:3}, left panel, shows $D$ as a function of $g_q/\kappa$ and $g_a/\kappa$ for a $^{133}$Cs atom, where $g_b/g_a = 1/\sqrt{45}$ \cite{Ostrowski_2022, Steck1998}.
			
\subsection{The cavity dark state}
\label{subsec:dark}
	
The system has also a non-zero probability to remain in an eigenstate of $\Big( g_q \sig_{q+}(\hat a + \hat b) + g_a \hat \sig_{a+} \hat a + g_b \hat \sig_{b+} \hat b + \text{H.c.} \Big)$, that is dark to the cavity, i.e., it is decoupled from the non-vacuum cavity states. This CDS $|CD\rangle$ can be expressed as
\be  \label{eq:CD}
	|CD\rangle = \mathcal{N} \Big( g_qg_b|G\rangle|+\rangle  -g_ag_b|E\rangle|g\rangle
	 + g_qg_a|G\rangle|-\rangle \Big)  |0\rangle_a|0\rangle_b,
\ee
with normalization constant
\be \label{eq:norm}
	\mathcal{N} = \Big( (g_ag_b)^2 + (g_qg_b)^2 + (g_qg_a)^2 \Big)^{-1/2}.
\ee
We assume here that the initial state is the normalized pure state $\ket{\bar \psi(0)} = |E,g,0,0 \rangle$,  $|\langle \bar \psi (0) | \bar \psi (0) \rangle|^2 =1$. Moreover, the system can fall into the CDS only if it does not emit a photon through the cavity. Hence, when the system is in the CDS, its initial norm is preserved. The probability for the system to fall into the CDS is thus given by the overlap of the initial state $| \psi (0) \rangle$ with $|CD\rangle$
		\bea \label{eq:P_CD}
			P_{CD} = & |\langle CD | \bar \psi(0) \rangle|^2\\
			       = & \frac{(g_ag_b)^2}{(g_ag_b)^2 + (g_qg_b)^2 + (g_qg_a)^2}.
		\eea
For non-zero spontaneous-emission rates of the QD and the atom, the $|CD\rangle$ state can, of course, decay. However,
our assumption of the Purcell regime guarantees that spontaneous decay of the dark state happens at a rate much smaller than the rate of coupling to the cavity (and the decay rate of the cavity itself).
In what follows, we take $\gamma_{\rm q,a,b}=0$ for simplicity, i.e. we assume that the CDS is stable, as can be seen by inserting Eq. \eqref{eq:CD} in Eq. \eqref{eq:proba_amp},
which gives $\dot Q = \dot A = \dot B = 0$. Additionally, it can be shown that
		\be
			P_a + P_b + P_{CD} = 1,
		\ee
meaning that the system can either emit a photon through the $a$ or $b$ mode or be trapped in the CDS.
Experimentally, we can know that the system is in the CDS if, after a sufficiently long time (in practice $t \gg 1/\kappa, t \ll 1/\gamma_{\rm q, a, b}$), no photon has been detected in the waveguide.

As discussed in~\cite{Ostrowski_2022}, if the amplitude of the dark state is relatively large, the photon emission rate will be reduced. If the goal of the experiment is to produce a directional single photon source as in~\cite{Ostrowski_2022}, suppression of the CDS amplitude is desirable. On the other hand, if the goal is to produce an entangled state of the atom and QD, as we discuss below, system parameters which produce a larger dark state amplitude may be set.
			
\section{Entanglement of the cavity dark state} 
\label{sec:ent}
		
The CDS exists in a subspace of the system decoupled from the cavity modes, where only the QD and atomic states can have non-zero amplitude.
In general the CDS has non-zero entanglement between the QD and the atom, which may be evaluated using two standard entanglement measures, the concurrence $C$ and the von Neumann entropy $S$.
Given that these measures are in one-to-one correspondence for low-dimensional pure states \cite{Wootters2001}, we choose to focus on the concurrence below.
		
\begin{figure}[tb!]
\centering
\includegraphics[width=\linewidth]{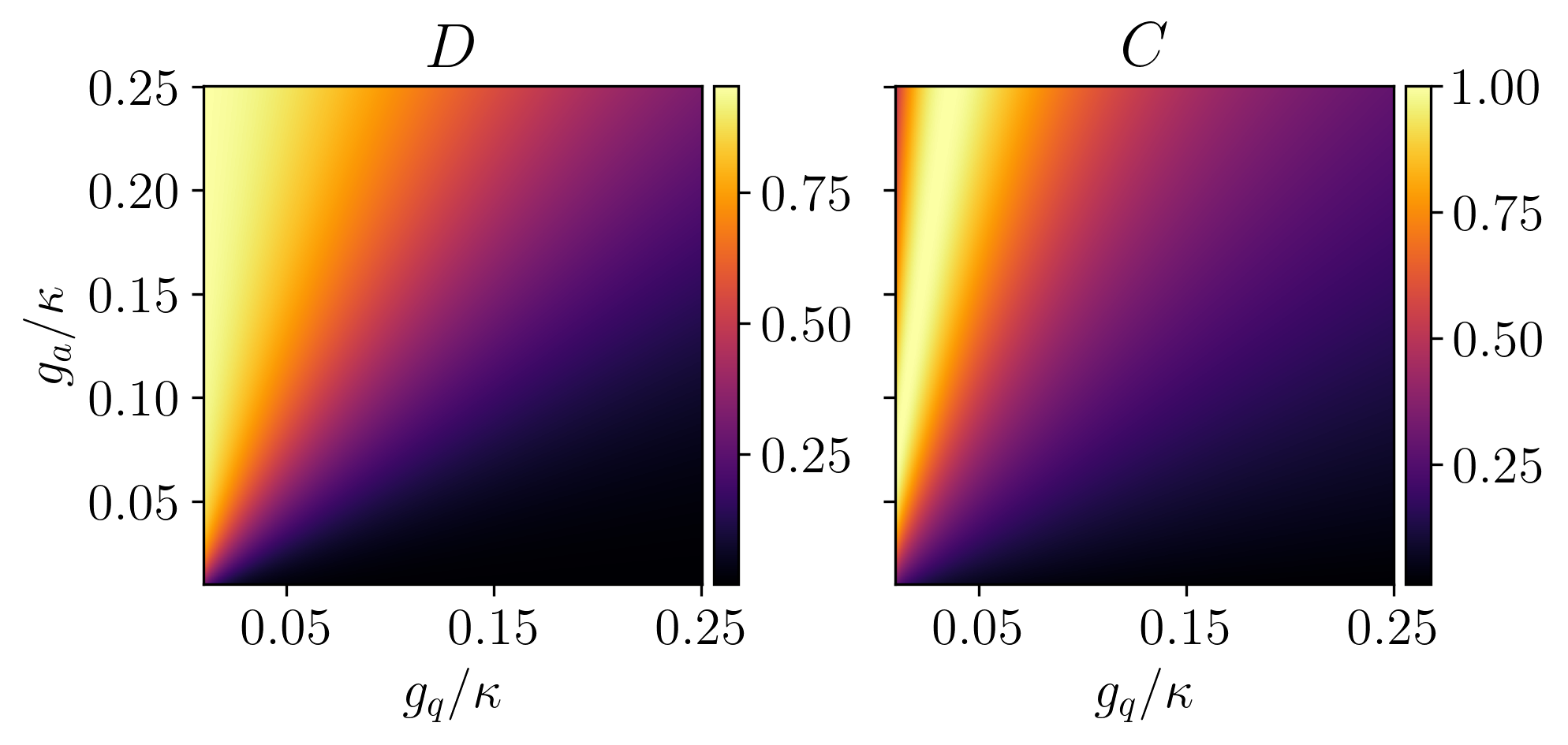}
\caption{(Left panel) Directionality, as defined in Eq. \eqref{eq:D_Gamma} in the parameter plane ($g_q/\kappa, g_a/\kappa$) for $g_b=g_a/\sqrt{45}$, with $\gamma_{q,a,b}=0$, $\Delta_{q,a,b}=0$.
(Right panel) Concurrence as defined in Eq. \eqref{eq:C_g}
plotted in the parameter plane ($g_q/\kappa, g_a/\kappa$) for $g_b=g_a/\sqrt{45}$, with $\gamma_{q,a,b}=0$, $\Delta_{q,a,b}=0$.}
\label{fig:3}
\end{figure}
		
\subsection{Calculation of concurrence}
			
We use a general definition of the concurrence (i.e. one which also applies to systems larger than two qubits) \cite{Milburn2002, Bhaskara_2017}
\be 
\label{eq:C_def}
C = \sqrt{2\Big(1 - \text{tr}(\hat \rho^2_\mathcal{M} ) \Big)},
\ee 
where $\hat \rho_\mathcal{M}$ is the reduced density matrix of the subsystem $\mathcal{M}$ after tracing out the complementary subsystem $\overline{\mathcal{M}}$.
We assume that the system is indeed in the CDS at the end of the evolution, so that it is described by a pure state of the form of Eq. \eqref{eq:CD}
\be
\label{eq:cdent}
|CD\rangle = Q|E\rangle|g\rangle + A|G\rangle|+\rangle + B|G\rangle|-\rangle,
\ee
expressed in the subspace of the QD and the atom, the cavity modes $|0\rangle_a|0\rangle_b$ being already factorized out. The corresponding density matrix $\hat\rho=|CD\rangle\langle CD|$ is
			\be \label{eq:rho_tot}
				\hat \rho =
				\begin{pmatrix}
					0 & 0   & 0   & 0   & 0 & 0 \\
					0 & A^2 & AB  & QA  & 0 & 0 \\
					0 & AB  & B^2 & QB  & 0 & 0 \\
					0 & QA  & QB  & Q^2 & 0 & 0 \\
					0 & 0   & 0   & 0   & 0 & 0 \\
					0 & 0   & 0   & 0   & 0 & 0
				\end{pmatrix},
			\ee 
written in the basis of the states $\{|G\rangle |g\rangle$, $|G\rangle |+\rangle$,  $|G\rangle |-\rangle$, $|E\rangle |g\rangle$, $|E\rangle |+\rangle$, $|E\rangle |-\rangle \}$. The amplitudes $A, B$, and $Q$ are all real for all times following Eqs. \eqref{eq:proba_amp}. This fact is of crucial importance in the coupling between the QD and the atom made explicit in \cite{Ostrowski_2022}.

Tracing out the atom gives			
\be \label{eq:rho_QD}
				\hat \rho_\text{QD} =
					\begin{pmatrix}
						A^2 + B^2 & 0 \\
						0 & Q^2
					\end{pmatrix},
\ee
which is the matrix of a mixed state (i.e. no coherence terms exist), with maximal mixedness when $Q^2=A^2+B^2 = 0.5$.

With this reduced state we can compute
\be \label{eq:C_CD}
				\text{tr}(\hat \rho_\text{QD}^2) = ( A^2 + B^2 )^2 + Q^4,
\ee
and obtain for the concurrence
\be \label{eq:C_first_expr}
				C = \sqrt{2\Bigg(1 - \Big(Q^4 + (A^2+B^2)^2 \Big) \Bigg)}.
\ee			
Now, using the normalization of $|CD\rangle$ from Eq. \eqref{eq:CD}, i.e., $|\langle CD|CD\rangle|^2 = Q^2+A^2+B^2=1$, we have
\be
\label{eq:finalC}
C = 2|Q|\sqrt{A^2+B^2}.
\ee
Note that tracing out the QD produces a $3\times3$ density matrix whose square has the same trace as Eq.~\ref{eq:C_CD}, and thus gives the same value of the concurrence.

Finally, we note that the concurrence may be given in terms of $g_q$, $g_a$ and $g_b$ by using Eq. (\ref{eq:CD}), i.e.,
$Q = -\mathcal{N}g_ag_b$, $A = \mathcal{N}g_qg_b$ and $B = \mathcal{N}g_qg_a$ with $\mathcal{N}$ defined in Eq. (\ref{eq:norm}).
This gives the following expression
\be \label{eq:C_g}
	C = \frac{2g_qg_ag_b \sqrt{ g_a^2 + g_b^2 }}{(g_ag_b)^2 + (g_qg_a)^2 + (g_qg_b)^2}.
\ee

 Let us now briefly check the extreme cases mentioned in the introduction.
 First, the case for $D\longrightarrow0$, which can be achieved in two ways. The intuitive way is by making the system symmetrical, as suggested in the introduction, i.e. $g_a=g_b=g$.
 Then, the cavity dark state becomes
 \begin{equation}
 |CD\rangle = \mathcal{N} g[g_q|G\rangle(|+\rangle+|-\rangle) - g|E\rangle|g\rangle],
 \end{equation}
 which, due to the assumed condition $g_q\ll g$ has vanishing entanglement. Specifically, the concurrence of this state is given by
 \begin{equation}
 C \approx 2\sqrt{2}\frac{g_q}{g}\ll 1.
 \end{equation}
 On the other hand, if we have an optically pumped atom in a state where $g_b\ll g_a$, $D$ can be made small irrespective of the value of $g_q$ by taking $g_a\longrightarrow 0$.
 In this case, $|CD\rangle\approx Ng_qg_a|G\rangle|-\rangle$, which is separable. As for the $D\longrightarrow 1$ case, this is achieved only when $g_b=0$, and it is trivial to see that $C=0$, in line with the non-existence of the dark state in that case.

 Regarding the interpretation of the above results, because the CDS is a sum of three orthogonal states, it inhabits a 3D subspace of the 6D atom-QD manifold.
 Given this, one would expect that the standard concurrence for 2 qubits could explain the entanglement in this case also.
 To formalize this intuition, we note that the entangled 2-qubit state
 \be \ket{\psi} = c_{10}|10\rangle + c_{01}|01\rangle\label{eq:qubit}\ee
 has a completely mixed reduced density matrix
 \be \label{eq:rho_qubit}
				\hat \rho_\text{R} =
					\begin{pmatrix}
						c_{10}^2& 0 \\
						0 & c_{01}^2
					\end{pmatrix},
\ee
for a partial trace over either qubit. The corresponding concurrence is $C=2|c_{10}c_{01}|$~\cite{Wootters2001}.
Hence, by identifying $c_{10}$ with $Q$ and $c_{01}$ with $\sqrt{A^2+B^2}$, we see that 
at the level of reduced density matrices the system is effectively described as a pair of qubits in a superposition like that given in Eq.~\eqref{eq:qubit}.

\section{Creation of CDS and measurement of entanglement}
\label{sec:exp}
In the experimental realization of the system shown in Fig.~\ref{fig:1}, the QD would typically be continuously driven by an external laser field. If no photons are detected
in the waveguide for a time $\sim T_g=1/g_q\ll1/\gamma_q$, it is assumed that the CDS has been prepared. Until then, the system will emit photons with a certain directionality,
which allows us to learn about the CDS entanglement as we shown below.

First, we note that both the directionality and concurrence are expressed in terms of the coupling constants $g_{q,a,b}$ in Eqs.~\eqref{eq:D_Gamma} and \eqref{eq:C_g} respectively. This allows us to relate the two quantities, measuring
the directionality of the series of photons emitted until the CDS is reached. From the directionality we can then determine the value of concurrence of the CDS. For a non driven QD, one would have to repeat the experiment many times to obtain good statistics of the photon emission and to determine the directionality to some desired precision. In the realistic case of continuous driving, the driving resets the system steadily and a typical cavity coupling rate of $g\sim 100$ MHz allows sufficient statistics to be gathered in less than 1s.

\subsection{Experimental considerations}
\label{sec:protocol}

In what follows, we will approximate the continuously driven system by considering ensembles of experiments prepared in the single excitation regime, i.e. in a pure state with the QD excited.
This allows us to use the analytical results we derived above, along with simple probabalistic arguments to determine the behavior of the system.

Let us define the ideal measurement protocol: First, the set of coupling constants is fixed by setting experimental parameters. Typically, the coupling of a solid state emitter on the surface of a circulating resonator, such as a silica bottle resonator~\cite{pollinger2010all}, micro toroid~\cite{aoki2006observation}, etc., will be decided by the resonator geometry and wavelength alone. For atoms trapped near to the resonator (using, e.g., a tweezer trap~\cite{thompson2013coupling,nayak2019real}), the distance from the resonator surface is a separately tunable parameter, which can be used to control the coupling. We note that an intriguing way to enhance the atom-resonator coupling if needed is to use $N$ atoms rather than one, in which case the collectively coupled ensemble has a coupling constant $\sqrt{N}g$, where $g$ is the single atom coupling constant, as demonstrated in a related setup recently~\cite{ruddell2017collective}. This would, in principle, result in entanglement between the QD and the entire ensemble, although detailed consideration of this system is beyond the scope of the current paper. We note that while the coupling of the QD is fixed after deposition, the coupling of the atom will depend on motion of the atom in the trap, stability of trap laser power and frequency, etc. We do not take detailed account of these issues here, but rather note the impressive advances in interfacing nanowaveguides and resonators in recent years as described in this paragraph and in the Introduction. It is of particular interest to note that cavity dark states of nanofiber coupled resonators have recently been successfully prepared and studied~\cite{kato2019observation,white2019cavity}.

After the system has been prepared, we perform the experiment $n_\text{runs}$ times. Each run starts with the same initial state, with the QD is in its excited state $|E\rangle$, the atom in its ground state $|g\rangle$, and no photon in the cavity. We then let the system evolve for a sufficiently long time ($t \gg 1/\kappa$). By detecting the photons in the waveguide, we count the number of photons emitted to the right $n_a$ and the left $n_b$.

If no photon is emitted, the system has fallen into the CDS and we can either use the so-prepared state, if sufficient statistics have been collected to characterize the entanglement to the desired precision, or ignore the run
and continue to make measurements of photons until $D$ is sufficiently well known.

Doing so we obtain $n_a+n_b = Pn_\text{runs}$ where $P$ is the probability of emitting a photon at all. We then estimate the directionality by
 \be 
 \label{eq:D_measured}
  D = \frac{n_b-n_a}{n_b+n_a}.
 \ee
 For $n_\text{runs} \longrightarrow \infty$, $D$ corresponds to its theoretical expression in Eq. \eqref{eq:D_def}.
 Using the relation from the following Section \ref{sec:con-dir}, we will deduce the value of the concurrence of the CDS produced with the current parameters (see Eqs. \eqref{eq:r_D-1} and \eqref{eq:C_r} below).
 
 The second step of our protocol would be necessary if measurements show that $C$ does not give the desired value.
 We then adjust the parameters, by changing the coupling strengths, and repeat step one from above.
 Once the measured value of directionality matched the requested value of the concurrence, we are sure that the system, once fallen into the CDS, is prepared in an entangled state, e.g., with maximal concurrence.
 
Our protocol depends on two conditions. One condition is a one-to-one relation of concurrence and directionality to be addressed in the next subsection. Another condition is the experimental stability of the system, e.g. with respect to laser drift, etc. Because photons are produced at a high rate (10s of MHz) compared to the typical drift time of experimental parameters ($\sim 1$ Hz), it should be possible in principle to meaningfully evaluate the directionality (and thus concurrence) for each parameter set.

\subsection{Correlation between concurrence and directionality}
\label{sec:con-dir}
	
The directionality $D$ and concurrence $C$ are expressed in terms of the coupling parameters $g_{q,a,b}$, see Eqs. \eqref{eq:D_Gamma} and \eqref{eq:C_g}. In the following we relate these two expressions. In a typical experimental setup, the QD comes with an unknown coupling constant $g_q$, while depending on the nature of the atom we know the ratio of its two internal transitions $r_a = g_b / g_a$, even if the single parameters $g_a$ and $g_b$. Hence, we
rewrite the above expressions as a function of the ratios $r=g_q/g_a$ and $r_a = g_b / g_a$:
\begin{eqnarray}
\label{eq:r_D-1}
 			D(r, r_a) & = & \frac{1-r_a^2}{1 + r_a^2 + 2r^2} \\
 \Longrightarrow \, r(D, r_a) &= &\sqrt{ \frac{(1-r_a^2) - (1+r_a^2)D}{2D} } \label{eq:r_D-2}
 ,
\end{eqnarray}
and 
\be 
\label{eq:C_r}
 	C(r, r_a) = \frac{2\sqrt{1+r_a^2}}{r/r_a + rr_a + r_a/r}.
 \ee
As shown in Fig. \ref{fig:3}, the concurrence $C$ is large in a narrow region $g_a \gg g_q$ that corresponds to a large directionality $D$. For a $^{133}$Cs atom with small $r_a=1/\sqrt{45}$, $C$ is maximal for small $r$, see Fig. \ref{fig:3}. The exact location of this region depends on the atomic ratio $r_a$.

\begin{figure}[tb] 
\centering
\includegraphics[width=\linewidth]{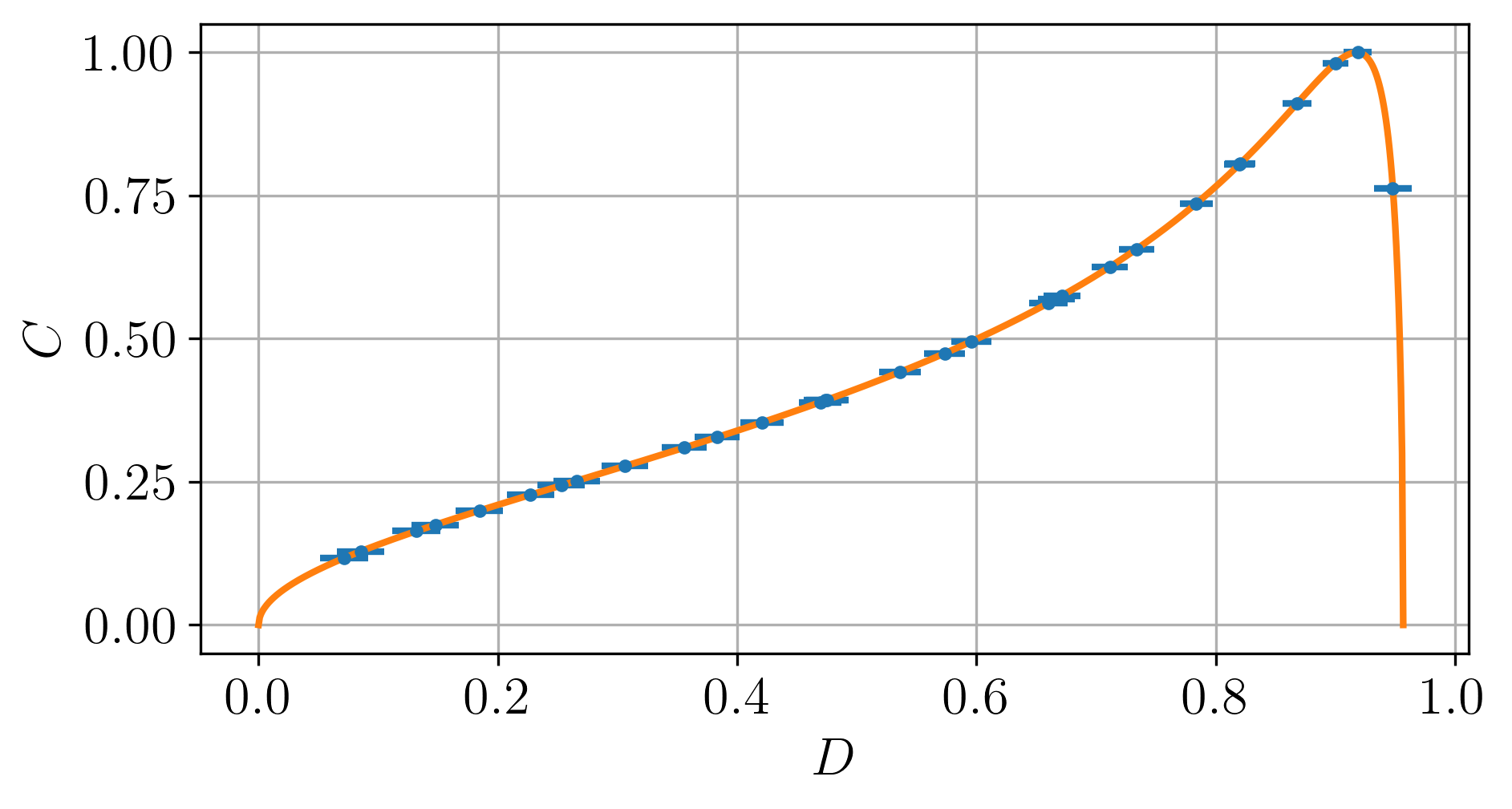}

\caption{Concurrence of the CDS vs. the directionality $D$ of the outgoing photons for the same coupling constants, for the case of $^{133}$Cs with $g_b/g_a = 1/\sqrt{45}$. The error bars indicate the standard deviation of $C$ obtained via the plotted relation assuming an error on the obtained value of $D$ with $n_\text{runs} = 5000$ of the protocol from sec. \ref{sec:protocol}.}

\label{fig:4}
\end{figure}

Using Eq. \eqref{eq:r_D-2}, $C$ can then formally be expressed as a function of $D$, through $r(D, r_a)$ for any fixed $r_a$. This is the correlation between the concurrence and the directionality anticipated above. We show this correlation graphically in Fig. \ref{fig:4}, again for the case of cesium with $r_a = g_b/g_a = 1/\sqrt{45}$. The value of $C$ increases with $D$ and reaches $C \approx 1$ for $D \approx 0.92$, before dropping rapidly above this value of $D$. For this case, the range of $D$ is bounded by $D_\text{max} \approx 0.95$. This maximally allowed value of $D$ depends on $r_a$, as explained in sec. \ref{subsec:dir}.

The steep drop off in $C$ as $D$ approaches 1 is due to the competing conditions for perfect concurrence and perfect directionality. Perfect concurrence occurs when i) a dark state exists and ii) $Q=\sqrt{A^2+B^2}=0.5$, while perfect directionality occurs in the case where i) interference blocks QD coupling to the $b$-mode and ii) the $|-\rangle$ level of the atom is completely decoupled and thus cannot populate the $b$-mode itself. However, the second condition
for perfect directionality conflicts with the first condition for perfect concurrence, since it requires the destruction of the CDS, which can only occur when the atom couples to \emph{both} cavity modes.
 
\subsection{Quantification of measurement errors}
\label{subsec:error}

Since the number of runs is finite in any experimental protocol (or equivalently, the total measurement time is finite for a driven QD), the measured directionality $D$ will have an uncertainty depending on the number of runs. This uncertainty propagates to the concurrence $C$, which was shown to depend on $D$ in the previous section. Since $C(D)$ drops dramatically for $D$ larger than $D\geq 0.95$, see Fig. \ref{fig:3}, this uncertainty is crucial in the determination of the concurrence.  Hence, a sufficient number of runs $n_\text{runs}$ is necessary to minimize the measurement error of $D$.

Building on the protocol suggested in Section~\ref{sec:protocol}, we quantify the error in the estimation of the concurrence $C$.
Given a set of coupling parameters $\{r, r_a\}$, we numerically simulate the system starting from the initial state assumed in sec. \ref{sec-2}. From the simulations we can estimate the probabilities $P_A$ and $P_B$ from Eq. \eqref{eq:proba-2}. We perform a simple statistical analysis based on these values by drawing $n_\text{runs}$ random numbers $D_i \in \{ -1, 1, 0\}$ with probabilities $P_A, P_B$ and $1-P_A-P_B$, respectively.
Note, that this is equivalent to simulating individual trajectories of the same system $n_\text{runs}$ times but is more efficient computationally. Discarding those values with $D_i = 0$, the mean of the data set can be used as the estimated value of the directionality, while the standard deviation divided by the number of non-zero $D_i$ values is the statistical uncertainty.
The latter translates into a statistical error on the extracted value of the concurrence via the relation \eqref{eq:C_r}. An example is shown in Fig. \ref{fig:4} for $n_\text{runs} = 5000$, for a broad range of $r$ values.

\begin{figure}[bt] 
	\centering
	\includegraphics[width=\linewidth]{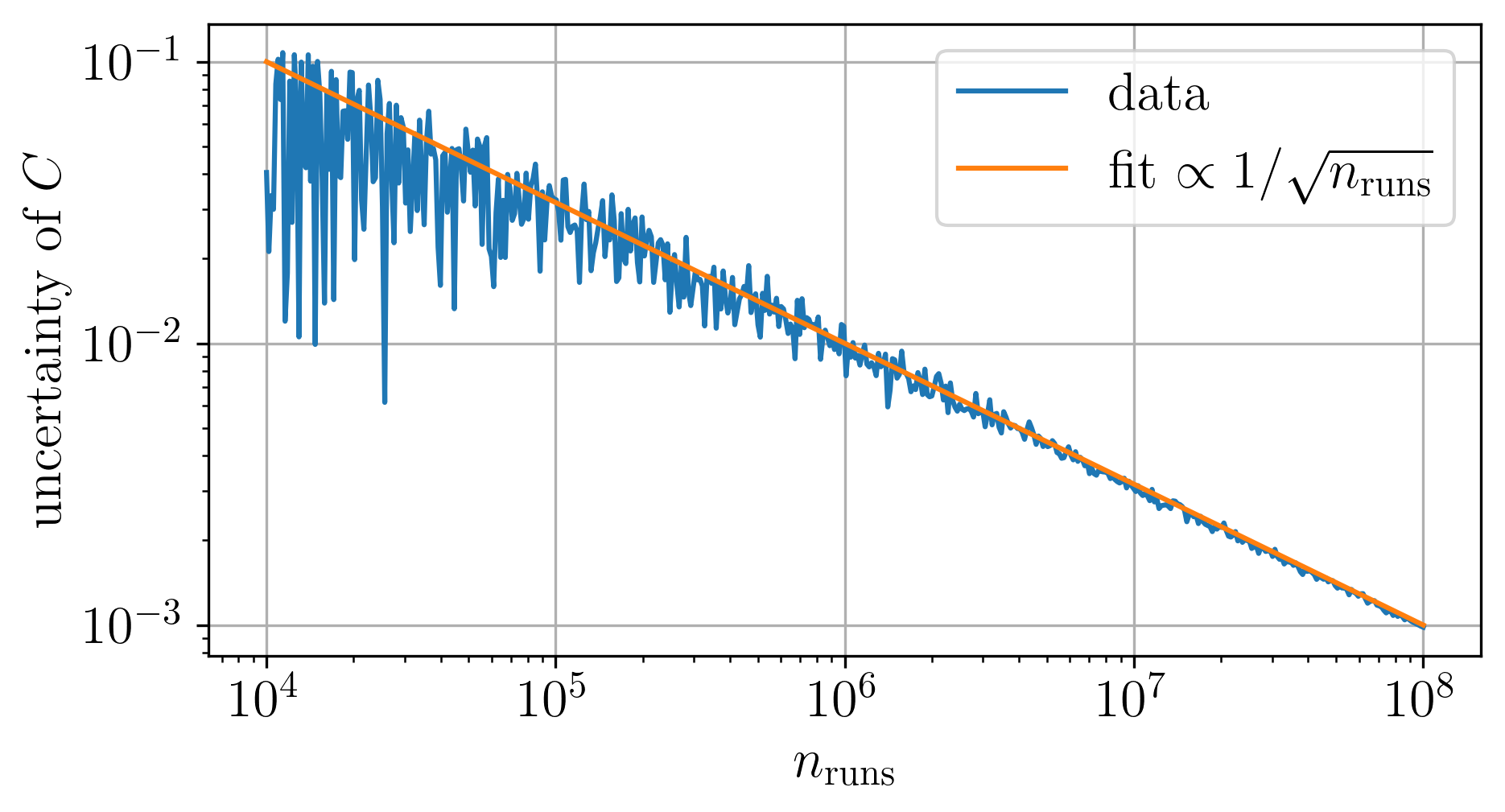}
	\caption{Uncertainty of the concurrence $C$ (blue line) around its peak $C \approx 1$ as a function of the number of runs $n_\text{runs}$. A fit $\propto 1/\sqrt{n_\text{runs}}$ (orange line) is added to show the rate of decay due to statistical fluctuations. The strong noisy behavior on the left comes from the fast drop of the curve $C(D)$ after its maximum, see Fig. \ref{fig:4}.} 
	\label{fig:5}
\end{figure}

Turning this around, one might also look at fixed $r$ varying $n_\text{runs}$. Assuming that we want to obtain a strongly entangled state, the most relevant point is to attain the peak with maximum concurrence $C \approx 1$. Tuning all parameters accordingly, $n_\text{runs}$ is scanned starting at the value $10000$ to reduce statistical fluctuations. The result is shown in Fig. \ref{fig:5} on a double-logarithmic scale. The uncertainty decreases with the expected $1/\sqrt{n_\text{runs}}$ scaling. The large "noise" seen in the interval up to $10^5$ stems from the stochastic nature of the process and indicates that $n_\text{runs}$ was not chosen sufficiently large, given the probabilities $P_A$ and $P_B$, for the result to converge properly. Since the typical emission rate of a Purcell enhanced QD is of the order of 100 MHz, repetitions with $n_\text{runs}=10^6-10^7$, are experimentally realistic. This would make the uncertainty on the estimate of the entanglement reasonably small $<10^{-2}$, see Fig. \ref{fig:5}.

\section{Discussion and Conclusion}
\label{sec:concl}

Extending the idea of ref. \cite{Ostrowski_2022}, where a quantum-dot emitter was coupled to an atom via a circular cavity in the bad cavity regime, we computed the directionality of emission and the entanglement of cavity dark states analytically using reasonable approximations. Our results allow us to relate the concurrence, which quantifies the entanglement between the quantum dot and the atom, with the directionality of the emitted photons.
We demonstrated that the concurrence of the dark state can be deduced from the experimentally observable directionality $D$ for the system before it settles into the CDS.
In particular, we found that concurrence obtains a theoretical maximum for large $D$, making the same system useful as both a highly directional single photon source, and a generator of quantifiable entanglement
between the atom and the solid state emitter.

Furthermore, the measurement of $D$ can be used to estimate the degree of entanglement to a high precision, even if we do not have direct knowledge of all the involved coupling strengths.
The same procedure as illustrated here for a generic quantum emitter and a cesium 133 atom can be applied to any other alkali atom as well.

Although the setup considered here is rather specialized, involving as it does a specific type of cavity and two different types of emitter, our analysis is likely to be applicable to more generic systems with only minor changes. Consider, for example, the case studied by White {\it et al.} in Ref.~\cite{white2019cavity}. In that study,  atoms were coupled to two independent linear cavities connected by an intervening cavity. This resulted in a dark state where only the intervening cavity could have non-zero photon population. In general, this state is a tripartite entangled state whose form is very similar to Eq.~\eqref{eq:CD} here. In Ref.~\cite{white2019cavity}, a difference in the coupling rates of the atoms to their respective cavities would be expected to lead to directional emission, and in the case of a hierarchy of cavity coupling strengths, similar to $g_q\ll g_a$ studied here, a regime where directionality and entanglement of the dark state are coupled might arise. Details of the relation between directionality and concurrence in such generic systems are beyond the scope of the present study, but suggest a fruitful avenue for future research.

Lastly, as mentioned in Section~\ref{sec:exp}, another area for future study is the entanglement of a QD with an atomic ensemble which may be more easy to realize experimentally, and could be treated using substantially
the same theoretical framework as the present study.

\medskip

\section*{Acknowledgements}

We are very grateful for support from Q-DYNAMO (EU HORIZON-MSCA-2022-SE-01) with project No. 101131418. S.W.  acknowledges funding from the National Recovery and Resilience Plan through Mission 4 Component 2 Investment 1.3, Call for tender No. 341 of 15/3/2022 of Italian MUR funded by NextGenerationEU, with project No. PE0000023, Concession Decree No. 1564 of 11/10/2022 adopted by MUR, CUP D93C22000940001, Project title "National Quantum Science and Technology Institute" (NQSTI), spoke 1. M.S. acknowledges funding from a JSPS KAKENHI Grant-in-Aid for Transformative Research Areas (Grant No. JP22H05135)

%
% ...

\bibliography{ref-dir}

\end{document}